\documentstyle[aps,pre]{revtex}

\begin{document}

\draft

\title{Generalized quantum Fokker-Planck, diffusion and Smoluchowski
equations with true probability distribution functions}

\author{Suman~Kumar~Banik$^1$, Bidhan~Chandra~Bag$^2$ and 
Deb~Shankar~Ray$^1$ }

\address{
$^1$Indian Association for the Cultivation of Science, Jadavpur, 
Calcutta 700 032, India.\\
$^2$Department of Chemistry, Visva-Bharati, Shantiniketan 731 235, India.}

\date{\today}

\maketitle

\begin{abstract}
%\scriptsize{
Traditionally, the quantum Brownian motion is described by  Fokker-Planck
or diffusion equations in terms of quasi-probability distribution functions,
e.g., Wigner functions. These often become singular
or negative in the full quantum regime. In this paper a
simple approach to non-Markovian theory of quantum Brownian motion 
using {\it true probability distribution functions} is 
presented. Based on an initial coherent state representation of the bath 
oscillators
and an equilibrium canonical distribution of the quantum mechanical 
mean values of their co-ordinates and momenta we derive a generalized quantum 
Langevin equation in $c$-numbers and show that the latter is amenable to a 
theoretical analysis in terms of the classical theory of non-Markovian 
dynamics. The corresponding Fokker-Planck, diffusion and the 
Smoluchowski equations
are the {\it exact} quantum analogues of their classical counterparts.
The present work is
{\it independent} of path integral techniques. The theory as 
developed here is a natural extension of its classical version and is
valid for arbitrary temperature and friction (Smoluchowski equation being
considered in the overdamped limit).
%}
\end{abstract}


%%%%%%%%%%%%%%%%%%%%%%%%%% INTRODUCTION %%%%%%%%%%%%%%%%%%%%%%%%%%%%%%%%%

\section{Introduction}

A model quantum system coupled to its environment forms the standard
paradigm of quantum Brownian motion. The initiation of early development
of this stochastic process took place around the middle of this century
\cite{whl,wlamb,lax}. A major impetus was 
the discovery of laser in sixties followed by significant
advancement in the field of quantum optics and laser physics in seventies
where the extensive applications of nonequilibrium quantum statistical
methods were made.
Various nonlinear optical processes/phenomena were
described with the help of operator Langevin equations, density operator 
methods and the associated quasi-classical distribution functions of 
Wigner, Glauber, Sudarshan and others centering around the quantum 
Markov processes \cite{whl,wlamb,lax,optics,dsr,tw}.
Subsequent to this early development the quantum theory of Brownian motion 
again emerged
as a subject of immense interest in early eighties when the problem of
macroscopic quantum tunneling was addressed by  
Leggett and others \cite{aoc,rmp-1,other-1,weiss,vhva} 
and almost simultaneously quantum Kramers' problem 
attracted serious attention of a number of workers 
\cite{pgw,rmp-2,prep,jrc}.
The method which received major appreciation
in eighties and nineties in the wide community of physicists and chemists 
in these studies is the real time functional integral \cite{path1,path2}.
This method has been shown to be an 
effective tool for treatment of quantum transition state \cite{qts}, 
dissipative quantum coherence effects \cite{rmp-1,dqc}
as well as incoherent quantum tunneling processes 
\cite{rmp-2,prep,iqtp} and many related problems \cite{recent}.

Inspite of this phenomenal success it may, however, noted that compared to
classical theory quantum theory of Brownian motion based on functional
integrals rests on a fundamentally different footing. While the classical 
theory is based on the differential equations for evolution of 
{\it true} probability density functions of the particle executing 
Brownian motion, the path integral methods rely on non-canonical quantization 
procedure and the evaluation of quantum partition function of the
particle interacting with the heat bath and one is, in general, led to the
time evolution equations of {\it quasi-probability distribution functions}
such as Wigner functions \cite{jrc,epw,dsr1,dsr2,dsr3,wf}.
The question is {\it whether
there is any natural extension of classical method to quantum domain
in terms of true probability distribution functions}. 
It is therefore worthwhile to seek {\it for a natural 
extension of the classical theory of Brownian motion to quantum domain} in 
the {\it non-Markovian regime} for {\it arbitrary friction} and 
{\it temperature} within the framework of a {\it well-behaved true
probabilistic} description.

\noindent
Our aim in this paper is thus twofold :

(i) to enquire whether there exists a quantum generalized Langevin equation
(QGLE) in $c$-numbers whose noise correlation satisfies the quantum
fluctuation-dissipation relation (FDR) but which (QGLE) at the same time
is a natural analogue of its classical counterpart.

(ii) to formulate the exact quantum Fokker-Planck and diffusion equations 
which are valid for arbitrary temperature and friction. We also intend to 
look for the overdamped limit to obtain the exact quantum analogue of
classical Smoluchowski equation.

%%%%%% New addition %%%%%%

Before proceeding further it is important to stress the motivation for the
present scheme:

(1) As we have already pointed out that the traditional theories of quantum
Brownian motion in optics \cite{whl,wlamb,lax,optics,dsr} and
condensed matter physics \cite{aoc}
are based on quasi-probability functions. Apart from their usual shortcomings
that they may become negative or singular \cite{rv}
in the full quantum regime
when the potential is nonlinear, the quasi-probability functions are, in
general, not valid for {\it non-Markovian processes} with {\it arbitrary}
noise correlation. While in majority of the quantum optical situations
Markovian description is sufficient, non-Markovian effects of noise
correlation are strongly felt in the problem of quantum dissipation in
condensed matter and chemical physics at low temperature. To include
these effects even in the case of a free particle [ see for example,
Ref.~\cite{vhva} ] one has to use a suitable cut-off frequency of the
heat bath to avoid intrinsic low frequency divergence. Clearly this
poses serious difficulties for studying transient behavior for
{\it arbitrary noise correlation} and {\it temperature}. In what follows
we show that the present treatment is free from such difficulties.

(2) Our second motivation is to understand {\it quantum-classical
correspondence} in the problem of Brownian motion in a transparent way.
To this end we note that in the classical theory the Fokker-Planck
equation with nonlinear potential contains derivatives of probability
distribution functions upto second order. The equations in terms of Wigner
functions on the other hand involve higher (than two) order derivatives
of distribution functions in the corresponding quantum formulation \cite{whz}.
The higher derivative terms contain powers of $\hbar$ and derivatives
of potential signifying purely quantum diffusion in which quantum corrections
and nonlinearity of the potential get entangled in the description of the
system. Because of the occurence of higher derivatives the positivity
of the distribution function is never ensured and the equation cannot
be treated as a quantum analogue of classical Fokker-Planck equation.
Any attempt to reduce the order of the derivatives to two
amounts to a semiclassical approximation. Again there exists no systematic
procedure for this reduction. Keeping in view of these problems we intend to
derive {\it exact} quantum analogues [ Eqs.(42), (47) and (50) ] of
classical Fokker-Planck, diffusion and Smoluchowski
equations, respectively, in terms of {\it true probability distribution
function} where the equations contain derivatives of distribution
functions upto second order only for which the diffusion coefficients are
positive definite. Since the equations are classical looking in form but
quantum mechanical in their content one can read the quantum drift and
diffusion coefficients and also construct the quantum corrections due to
the nonlinearity of the system systematically order by order
in a straightforward way so that quantum-classical
correspondance can be checked simply by taking limit $\hbar \rightarrow 0$
both in Markovian and in non-Markovian description. We mention in passing
that in contrast to a recent treatment \cite{apg} of large friction limit 
in a similar context, the quantum Smoluchowski equation as discussed here
retains its validity in the full quantum regime as $T \rightarrow 0$.

(3) Since over the last two decades classical non-Markovian theories 
\cite{mazo,adelman} and
numerical methods of generating classical noise processes have made
a significant progress \cite{num1,num2,num3},
the mapping of quantum theory of Brownian motion
into a classical form, as achieved here, suggests that the classical
treatment can be extended to quantum domain without much difficulty.
Since the present scheme describes the generation of quantum noise
[ Eqs.~(10) and (11) ]
as classical numbers which follow quantum fluctuation-dissipation relation
it is easy to comprehend that the classical numerical techniques of
generation of noise and solving stochastic Langevin equation
\cite{num1,num2,num3}
can be utilized in the present case in a straightforward way to solve
quantum Langevin equation \cite{dbsd2}. The procedure is therefore much
easy to implement compared to other methods like path integral Monte Carlo
techniques \cite{makri}.

%%%%%% New addition %%%%%%

In what follows we consider the standard system-reservoir model and
make use of the coherent state representation of 
the bath oscillators to derive a GLE for quantum
mechanical mean value of position of a particle in contact with a thermal 
bath whose quantum mechanical properties can be defined in terms of a 
classical-looking noise term and a canonical distribution of initial
quantum mechanical mean values of the co-ordinates and momenta of the bath.
This simple approach allows us to show that although the equation is
essentially quantum mechanical it is amenable to a theoretical
analysis in terms of the classical theory of non-Markovian dynamics
\cite{mazo,adelman}.

The rest of the paper is organized as follows :  The system reservoir model, 
the associated QGLE and the canonical distribution for the bath oscillators
have been introduced in Sec.~{II}. This is followed by a general analysis of
QGLE in Sec.~{III} and an illustration with an exponential memory kernel
in Sec.~{IV} to calculate the variances required for setting up a quantum
Fokker-Planck equation and a quantum diffusion equation in sections
V and VI, repectively. Section VII is devoted to quantum overdamped limit
and Smoluchowski equation. The paper is summarized and concluded in
Sec.~{VIII}.

%%%%%%%%%%%%%%%%%%%%% SECTION-II %%%%%%%%%%%%%%%%%%%%%%%%%%%%%%%%%%%%%

\section{The quantum Generalized Langevin equation (QGLE) in $c$-numbers }

We consider a particle in a medium. The latter is modeled as a set of
harmonic oscillators with frequency $\{ \omega_i \}$. Evolution of such a 
quantum open system has been studied over the last several decades under a 
variety of reasonable assumptions. Specifically our interest here is to
develop an exact 
description of quantum Brownian motion within the perview of this model
described by the following Hamiltonian \cite{rz},

\begin{equation}
\label{eq1}
\hat{H} = \frac{ \hat{p}^2 }{2} + V ( \hat{x} ) 
+ \sum_j \left [  \; \frac{ \hat{p}_j^2 }{2}
+ \frac{1}{2} \kappa_j ( \hat{q}_j - \hat{x} )^2 \; \right ] \; \; .
\end{equation}

\noindent
Here $\hat{x}$ and $\hat{p}$ are co-ordinate and momentum operators of
the particle and the set $ \{ \hat{q}_j,\hat{p}_j \}$ is the set of 
co-ordinate and momentum operators for the reservoir oscillators coupled 
linearly to the system through their coupling coefficients $\kappa_j$. 
The potential $V(\hat{x})$ is due to the external force field for the 
Brownian particle. The co-ordinate and momentum operators follow the
usual commutation relation [$\hat{x}, \hat{p}$] = $i\hbar$ and
[$\hat{q}_j, \hat{p}_j$] = $i\hbar \delta_{ij}$.
Note that in writing down the Hamiltonian no rotating 
wave approximation has been used.

Eliminating the reservoir degrees of freedom in the usual way 
\cite{whl,west,jkb,flo} we obtain the
operator Langevin equation for the particle,
\begin{equation}
\label{eq2}
\ddot{ \hat{x} } (t) + \int_0^t dt' \; \gamma (t-t') \; \dot{ \hat{x} } (t')
+ V' ( \hat{x} )
= \hat{F} (t) \; \; ,
\end{equation}

\noindent
where the noise operator $\hat{F} (t)$ and the memory kernel 
$\gamma (t)$ are given by
\begin{equation}
\label{eq3}
\hat{F} (t) = \sum_j \left [ \; 
\left \{ \hat{q}_j (0) - \hat{x} (0) \right \} \;
\kappa_j \; \cos \omega_j t +
\hat{p}_j (0) \; \kappa_j^{1/2} \; \sin \omega_j t \; \right ]
\end{equation}

\noindent
and
\begin{equation}
\label{eq4}
\gamma (t) = \sum_j \kappa_j \; \cos \omega_j t \; \; ,
\end{equation}

\noindent
with $\kappa_j = \omega_j^2$ ( masses have been assumed to be unity ). 

The Eq.(\ref{eq2}) is an exact quantized operator Langevin equation which is
now a standard textbook material \cite{whl,optics} and for which
the noise properties
of $\hat{F} (t)$ can be defined using a suitable initial canonical 
distribution of the bath co-ordinates and momenta. Our aim here is 
to replace it by an equivalent QGLE in $c$-numbers. Again this is not a new 
problem so long as one is restricted to standard quasi-probabilistic methods 
using, for example, Wigner functions \cite{jrc,epw,dsr1,dsr2,dsr3,wf}. 
To address the problem of quantum non-Markovian dynamics in terms of a 
{\it true probabilistic description} 
we, however, follow a different procedure. 
We {\it first} carry out the 
{\it quantum mechanical average} of Eq.(\ref{eq2})
\begin{equation}
\label{eq8}
\langle \ddot{ \hat{x} } (t) \rangle + 
\int_0^t dt' \; \gamma (t-t') \; \langle \dot{ \hat{x} } (t') \rangle
+ \langle V' ( \hat{x} ) \rangle
= \langle \hat{F} (t) \rangle
\end{equation}

\noindent
where the average $\langle \ldots \rangle$
is taken over the initial product separable quantum states
of the particle and the bath oscillators at $t=0$,
$| \phi \rangle \{ | \alpha_1 \rangle | \alpha_2 \rangle \ldots
| \alpha_N \rangle \} $.
Here $| \phi \rangle$ denotes any arbitrary 
initial state of the particle and
$| \alpha_i \rangle$ corresponds to the initial
coherent state of the $i$-th bath oscillator. $|\alpha_i \rangle$
is given by 
$|\alpha_i \rangle = \exp(-|\alpha_i|^2/2) 
\sum_{n_i=0}^\infty (\alpha_i^{n_i} /\sqrt{n_i !} ) | n_i \rangle $,
$\alpha_i$ being expressed in terms of the mean values of the co-ordinate
and momentum of the $i$-th oscillator,
$\langle \hat{q}_i (0) \rangle = ( \sqrt{\hbar} /2\omega_i)
(\alpha_i + \alpha_i^\star )$ and
$\langle \hat{p}_i (0) \rangle = i \sqrt{\hbar\omega_i/2 }
(\alpha_i^\star - \alpha_i )$, respectively.
It is important to note that $\langle \hat{F} (t) \rangle$
of Eq.(\ref{eq8}) is a classical-like noise term which, in general, is a
non-zero number because of the quantum mechanical averaging over the 
co-ordinate and momentum operators of the bath oscillators with respect to 
the initial coherent states and arbitrary initial state of the particle
and is given by
\begin{equation}
\label{eq9}
\langle \hat{F} (t) \rangle = \sum_j \left [ \; 
\left \{  \langle \hat{q}_j (0) \rangle - \langle \hat{x} (0) \rangle 
\right \} \; \kappa_j \; \cos \omega_j t +
\langle \hat{p}_j (0) \rangle \; \kappa_j^{1/2} \; \sin \omega_j t 
\; \right ] \; \; .
\end{equation}

\noindent
It is convenient to rewrite the $c$-number equation (\ref{eq8})
as follows ;
\begin{equation}
\label{eq10}
\langle \ddot{ \hat{x} } (t) \rangle + 
\int_0^t dt' \; \gamma (t-t') \; \langle \dot{ \hat{x} } (t') \rangle
+ \langle V' ( \hat{x} ) \rangle
= F (t)
\end{equation}

\noindent
where we let the quantum mechanical mean value 
$ \langle \hat{F} (t) \rangle = F (t)$. We now turn to the {\it second}
averaging. To realize $F(t)$ as an effective 
$c$-number noise we now assume that the momenta 
$\langle \hat{p}_j (0) \rangle$ and the shifted co-ordinates
$\{ \langle \hat{q}_j (0) \rangle - \langle \hat{x} (0) \rangle \}$
of the bath oscillators are distributed according to a canonical distribution
of Gaussian forms as
\begin{equation}
\label{eq13}
{\cal P}_j = {\cal N} \; 
\exp \left \{ \frac{ - \; [ \langle \hat{p}_j (0) \rangle^2 +
\kappa_j \left \{
\langle \hat{q}_j (0) \rangle - \langle \hat{x} (0) \rangle \right \}^2  
] }{
2 \hbar \omega_j \left ( \bar{n}_j + \frac{1}{2} \right ) }
\right \} 
\end{equation}

\noindent
so that for any quantum mechanical mean value
$O_j ( \langle\hat{p}_j (0) \rangle, 
\{ \langle \hat{q}_j (0) \rangle  - \langle \hat{x} (0) \rangle \} )$ the 
statistical average $\langle \ldots \rangle_S$ is
\begin{eqnarray}
\label{eq14}
\langle O_j \rangle_S & = & \int 
O_j ( \langle \hat{p}_j (0) \rangle, 
\{ \langle \hat{q}_j (0) \rangle  - \langle \hat{x} (0) \rangle \} )
\nonumber \\
& & \times \; {\cal P}_j ( \langle \hat{p}_j (0) \rangle, 
\{ \langle \hat{q}_j (0) \rangle - \langle \hat{x} (0) \rangle \} ) \;
d\langle \hat{p}_j (0) \rangle \;  
d \{ \langle \hat{q}_j (0) \rangle - \langle \hat{x} (0) \rangle \} \; \; .
\end{eqnarray}

\noindent
Here $\bar{n}_j$ indicates the average thermal photon number of the $j$-th
oscillator at temperature $T$ and 
$\bar{n}_j = 1/[\exp \left ( \hbar \omega_j/k_BT \right ) - 1]$ and
${\cal N}$ is the normalization constant.

The distribution (\ref{eq13}) and the definition of statistical average
(\ref{eq14}) imply that $F(t)$ must satisfy
\begin{equation}
\label{eq11}
\langle F (t) \rangle_S = 0
\end{equation}

\noindent
and
\begin{equation}
\label{eq12}
\langle F (t) F (t') \rangle_S
= \frac{1}{2} \sum_j \kappa_j \; \hbar \omega_j \; 
\left ( \coth \frac{\hbar\omega_j}{2k_BT} \right ) \; \cos \omega_j (t-t')
\; \; .
\end{equation}

\noindent
That is, the $c$-number noise $F(t)$ is such that it is zero centered and 
satisfies the standard quantum fluctuation-dissipation relation (FDR)
as known in the literature \cite{west} in
terms of quantum statistical average of the noise operators. 

To proceed further we now add the force term $V'(\langle \hat{x} \rangle )$
on both sides of Eq.(\ref{eq10}) and rearrange it to obtain formally
\begin{equation}
\label{eqn1}
\ddot{X} (t) + \int_0^t dt' \; \gamma (t-t') \; \dot{X} (t')
+ V' ( X )
= F (t) + Q (X, t)
\end{equation}

\noindent
where we let $\langle \hat{x} (t) \rangle = X (t)$ for simple notational
convenience and
\begin{equation}
\label{eqn2}
Q (X, t) = V' ( \langle \hat{x} \rangle ) - \langle V' (\hat{x} ) \rangle
\end{equation}

\noindent
represents the quantum mechanical dispersion of the force operator
$V'(\hat{x})$ due to the system degree of freedom. Since $Q(t)$ is a 
quantum fluctuation term Eq.(\ref{eqn1}) offers a simple interpretation.
This implies that the classical looking QGLE is governed by a $c$-number
quantum noise $F(t)$ which originates from the quantum mechanical
heat bath characterized by the properties (\ref{eq11}) and
(\ref{eq12}) and a quantum
fluctuation term $Q(t)$ due to the quantum nature of the system 
characteristic of the nonlinearity of the potential. In Sec.~{VII}
we give a recipe for calculation of $Q(t)$.

Summarizing the above discussions we point out that it is possible to 
formulate a QGLE (\ref{eqn1}) of the
quantum mechanical mean value of position of
a particle in a medium, provided the classical-like noise term $F(t)$
satisfies (\ref{eq11}) and (\ref{eq12}) where the ensemble average has to
be carried out with the distribution (\ref{eq13}). It is thus apparent
that to realize $F(t)$ as a noise term we have split up the standard
quantum statistical averaging procedure
into a quantum mechanical mean $\langle \ldots \rangle$ 
by explicitly using an initial coherent state representation of the bath 
oscillators and then a statistical average $\langle \ldots \rangle_S$
of the quantum mechanical mean values. 
Two pertinent points are to be noted :  First, it may be easily verified that
the distribution of quantum mechanical
mean values of the bath oscillators (\ref{eq13})
reduces to classical Maxwell-Boltzmann
distribution in the thermal limit, $\hbar \omega_j \ll k_BT$.
Second, the vacuum term in the distribution (\ref{eq13}) prevents the 
distribution of quantum mechanical mean values from being singular at
$T=0$ ; or in other words the width of distribution remains finite even 
at absolute zero, which is a simple consequence of uncertainty principle.

%%%%%%%%%%%%%%%%%%%%%%%%%%% SECTION - III %%%%%%%%%%%%%%%%%%%%%%%%%%%%%%%%%

\section{General analysis: Damped free particle
\label{sec3}
}

It is now convenient to rewrite QGLE (\ref{eqn1}) of quantum mechanical mean 
value of position of a particle in the absence of any external force field
in the form
\begin{equation}
\label{eq15}
\ddot{X} (t) + \int_0^t \gamma (t-t') \; \dot{X} (t') \; dt' = F(t)
\end{equation}

\noindent
$\gamma (t)$ is the dissipative memory kernel as given by 
Eq.(\ref{eq4}) and $F(t)$ is the zero centered stationary noise, i.e.,
\begin{equation}
\label{eq16}
\langle F(t) \rangle_S = 0 \; \; {\rm and} \; \;
\langle F(t) F(t') \rangle_S = C (|t-t'|) = C(\tau) 
\end{equation}

\noindent
where $C(t)$ is the correlation function which in the equilibrium state
is connected to the memory kernel $\gamma (t)$ through FDR of the form
\cite{aoc}
\begin{equation}
\label{eq17}
C (t-t') = \frac{1}{2} \int_0^\infty d\omega \; \kappa (\omega) 
\varrho (\omega) \; \hbar \omega \; \left ( \coth 
\frac{\hbar \omega}{2k_BT} \right ) \; \cos \omega (t-t')
\end{equation}

\noindent
Eq.(\ref{eq17}) is the continuum version of Eq.(\ref{eq12}). 
$\varrho (\omega)$ denotes the density of modes of the bath oscillators.
Here it is important to note that Eq.(\ref{eq17}) is the generalized FDR
valid at any arbitrary temperature $T$. $\gamma (t-t')$ is the continuum
version of Eq.(\ref{eq4}) and is given by
\begin{equation}
\label{eq18}
\gamma (t-t') = \int_0^\infty d\omega \; \kappa (\omega) \varrho (\omega) \;
\cos \omega (t-t') \; \; .
\end{equation}

\noindent
In the high temperature limit, i.e.,
for $\hbar\omega \ll k_BT$ we arrive at the wellknown classical FDR of the 
second kind \cite{kubo}
\begin{equation}
\label{eq19}
C (t-t') = k_BT \; \gamma (t-t') \; \; .
\end{equation}

The general solution of Eq.(\ref{eq15}) is given by
\begin{equation}
\label{eq20}
X (t) = \langle X (t) \rangle_S + \int_0^t H (t-\tau) \; F(\tau) \; d\tau
\end{equation}

\noindent
where
\begin{equation}
\label{eq21}
\langle X (t) \rangle_S = X_0 + V_0 H(t)
\end{equation}

\noindent
with $X_0 = X (0)$ and $V_0 = \dot{X} (0)$ being the initial quantum
mechanical mean values of position and velocity of the particle,
respectively. $H(t)$ is the inverse form of the Laplace transform
\begin{equation}
\label{eq22}
\tilde{H} (s) = \frac{1}{s^2 +s \tilde{\gamma} (s) }
\end{equation}

\noindent
with
\begin{equation}
\label{eq23}
\tilde{\gamma} (s) = \int_0^\infty \gamma (t) e^{-st} dt
\end{equation}

\noindent
is the Laplace transform of dissipative memory kernel $\gamma (t)$. The time
derivative of Eq.(\ref{eq20}) gives
\begin{equation}
\label{eq24}
V (t) = \langle V (t) \rangle_S + \int_0^t h(t-\tau) \; F(\tau) \; d\tau
\end{equation}

\noindent
where
\begin{equation}
\label{eq25}
\langle V (t) \rangle_S = V_0 h(t)
\end{equation}

\noindent
and
\begin{equation}
\label{eq26}
h (t) = \frac{d H(t)}{dt} \; \; .
\end{equation}

\noindent
Hence
\begin{equation}
\label{eq27}
\tilde{h} (s) = \frac{1}{s+\tilde{\gamma} (s)} \; \; .
\end{equation}

Before proceeding further it is important to recall the physical significance
of the two function $H(t)$ and $h(t)$. It has already been assumed that
the initial quantum mechanical
velocity $V_0$ is independent of the random force $F(t)$,
\begin{equation}
\label{eq28}
\langle V_0 F (t) \rangle_S = 0 \; \; .
\end{equation}

\noindent
Thus multiplying Eqs. (\ref{eq20}) and (\ref{eq24}) by $V_0$ and using 
relation (\ref{eq28}) we obtain,
\begin{equation}
\label{eq29}
\langle V_0 V (t) \rangle_S/\langle V_0^2 \rangle_S = h (t) \; \; ,
\end{equation}

\begin{equation}
\label{eq30}
\langle V_0 (X (t) - X_0) \rangle_S / \langle V_0^2 \rangle_S = H(t) \; \; .
\end{equation}

\noindent
Hence $H(t)$ and $h(t)$ are the two relaxation functions ; $h(t)$ measures
how the quantum mechanical mean velocity
forgets its initial value and $H(t)$ measures how the quantum
mechanical mean displacement forgets the initial velocity. 
As a result quantum mechanical mean velocity of the 
particle relaxes to a stationary state with zero statistical 
average of the quantum mechanical mean velocity.

Now using the symmetry property of the correlation function
\begin{eqnarray*}
\langle F (t) F (t') \rangle_S = C (t-t') = C (t'-t)
\end{eqnarray*}

\noindent
and using the solution for $X(t)$ and $V(t)$ we obtain the following
expressions of the variances,

\begin{mathletters}

\begin{eqnarray}
\sigma_{XX}^2 (t) & \equiv & 
\langle [ X (t) - \langle X (t) \rangle_S ]^2 \rangle_S \nonumber \\
& = & 2 \int_0^t H(t_1) \; dt_1 \int_0^{t_1} H(t_2) \; C (t_1 - t_2)
dt_2 \; \; , \label{eq31a} \\
\sigma_{VV}^2 (t) & \equiv & 
\langle [ V (t) - \langle V (t) \rangle_S ]^2 \rangle_S \nonumber \\
& = & 2 \int_0^t h(t_1) \; dt_1 \int_0^{t_1} h(t_2) \; C (t_1 - t_2)
dt_2 \; \; {\rm and} \label{eq31b} \\
\sigma_{XV}^2 (t) & \equiv & 
\langle [ X (t) - \langle X (t) \rangle_S ] 
[ V (t) - \langle V (t) \rangle_S ] \rangle_S = \frac{1}{2} 
\dot{\sigma}_{XX}^2 (t) \nonumber \\
& = & \int_0^t H(t_1) \; dt_1 \int_0^t h(t_2) \; C (t_1 - t_2)
dt_2 \; \; . \label{eq31c} 
\end{eqnarray}

\end{mathletters}

\noindent
The above three expressions are valid for arbitrary temperature 
and friction and include quantum effects. 
However in the high temperature classical
limit ( i.e., $\hbar \omega \ll k_BT$ ) one can derive 
simplified versions of the variances
\begin{mathletters}
\begin{eqnarray}
\sigma_{XX}^2 (t) & = & k_BT \left [ 2 \int_0^t H(t') dt' - H^2 (t)  
\right ] \; \; ,\\
\sigma_{VV}^2 (t) & = & k_BT \left [ 1 - h^2 (t) \right ] \; \; {\rm and} \\
\sigma_{XV}^2 (t) & = &  k_BT \; H(t)[ 1 - h(t) ] \; \; .
\end{eqnarray}
\end{mathletters}

Before closing this section we emphasize a pertinent point at this stage.
The (\ref{eq31a})-(\ref{eq31c}) are the expressions for statistical
variances of the quantum mechanical mean values $X$ and $V$. These are not
to be confused with the standard quantum mechanical variances which are
connected through uncertainty relations.

%%%%%%%%%%%%%%%%%%%%%%%%%%%%% SECTION IV %%%%%%%%%%%%%%%%%%%%%%%%%%%%%%%%%%

\section{A specific example : exponentially correlated memory kernel }

The very structure of $\gamma (t)$ given in Eq.(\ref{eq18}) 
suggests that it is quite general
and a further calculation requires a prior knowledge of the density of modes
$\varrho (\omega)$ of the bath oscillators. As an specific case we consider
in the continuum limit,
\begin{equation}
\label{eq33}
\kappa (\omega) \varrho (\omega) = \frac{2}{\pi} \; 
\frac{\gamma_0}{1+\omega^2 \tau_c^2} 
\end{equation}

\noindent
so that $\gamma (t)$ takes the wellknown form,
\begin{equation}
\label{eq34}
\gamma (t) = \frac{\gamma_0}{\tau_c} e^{-|t|/\tau_c} \; \; ,
\end{equation}

\noindent
where $\gamma_0$ is the damping constant and $\tau_c$ refers to correlation
time of the noise.
Once we get an explicit expression of $\gamma (t)$ in closed form and its
Laplace transform, it is possible to make use of Eq.(\ref{eq22}) to
calculate the relaxation function $H(t)$, which for the present case is 
given by
\begin{equation}
\label{eq35}
H (t) = \frac{1}{\gamma_0} \left [ 1 - {\cal A} e^{-t/2\tau_c} 
\sin ( \lambda t + \alpha) \right ]
\end{equation}

\noindent
where
\begin{equation}
\label{eq36}
{\cal A} = \frac{\gamma_0}{\lambda} \; \; , \; \;
\lambda = \left ( \frac{\gamma_0}{\tau_c} - \frac{1}{4\tau_c^2} 
\right )^{1/2} \; \; {\rm and} \; \;
\alpha = \tan^{-1} \left ( 
\frac{2\lambda \tau_c}{1-2\gamma_0 \tau_c} \right ) \; \; .
\end{equation}

Now making use of the expressions for $H(t)$ and 
the correlation function $C(t)$
in Eqs.(\ref{eq31a})-(\ref{eq31c}) we calculate explicitly after a long
but straightforward algebra the time dependent expressions of the variances
of the quantum mechanical mean value of position and momentum of the 
particle,
\begin{equation}
\label{eq37}
\sigma_{XX}^2 (t) = \frac{2 \hbar}{\pi} \int_0^\infty
\frac{\omega}{1+\omega^2 \tau_c^2} \; \left ( 
\coth \frac{\hbar \omega}{2k_BT} \right ) \; {\cal F}_X (\omega, t) \;
d\omega
\end{equation}

\begin{equation}
\label{eq38}
\sigma_{VV}^2 (t) = \frac{2 \gamma_0 \hbar}{\pi \lambda^2} \int_0^\infty
\frac{\omega}{1+\omega^2 \tau_c^2} \; \left ( 
\coth \frac{\hbar \omega}{2k_BT} \right ) \; {\cal F}_V (\omega, t) \;
d\omega
\end{equation}

\noindent
and
\begin{equation}
\label{eq39}
\sigma_{XV}^2 (t) = \frac{1}{2} \dot{\sigma}_{XX}^2 (t)
\end{equation}

\noindent
In Appendix-A we provide the explicit structures of 
${\cal F}_X (\omega, t)$ and ${\cal F}_V (\omega, t)$.

To examine the consistency of our calculation we check long time behaviour
of the classical high temperature Ohmic limit of the variances
$\sigma_{XX}^2 (t)$ and $\sigma_{VV}^2 (t)$. In this limit we have
\begin{eqnarray*}
\sigma_{XX}^2 (t) = \frac{4k_BT}{\pi} \int_0^\infty d\omega \;
\frac{1}{1+\omega^2 \tau_c^2} \; {\cal F}_X (\omega , t)
\end{eqnarray*}

\noindent
Only the first term of ${\cal F}_X (\omega , t)$ gives the long time
behaviour of $\sigma_{XX}^2 (t)$ in the Markovian limit, 
contribution of the rest of the terms being zero.
Taking this leading order contribution we have
\begin{eqnarray*}
\sigma_{XX}^2 (t) & = & \frac{4k_BT}{\pi \gamma_0} \int_0^\infty d\omega \;
\frac{1}{1+\omega^2 \tau_c^2} \; \frac{1}{\omega^2} (1-\cos \omega t)
\nonumber \\
& = & \frac{8k_BT}{\pi \gamma_0} \; 
\left ( \left. \frac{1}{1+\omega^2 \tau_c^2} \right |_{\omega=0} \right )
\int_0^\infty d\omega \; \frac{\sin^2 \frac{1}{2} \omega t}{\omega^2}
\end{eqnarray*}

\noindent
which gives
\begin{equation}
\label{eq40}
\sigma_{XX}^2 (t) = \frac{2k_BT}{\gamma_0} t \; \; \; {\rm for} \; 
t \rightarrow \infty \; \; .
\end{equation}

\noindent
Similarly one can show that for classical high temperature 
Markovian limit
\begin{equation}
\label{eq41}
\sigma_{VV}^2 (t) = k_BT \; \; \; {\rm for} \; 
t \rightarrow \infty \; \; .
\end{equation}

Since we are unable to evaluate analytically further
the explicit time dependent structures of the variances in the general case,
we take resort to numerical integration of Eqs.(\ref{eq37}) and (\ref{eq38}).
In Figs.(1) and (2) we show the short time and long time behaviour of the
variances $\sigma_{XX}^2 (t)$ as functions of time for different values
of temperature but for a fixed value of correlation time, $\tau_c$. It is
apparent that while the short time dynamics has a simple $t^2$ behaviour,
asymptotic dependence is linear in $t$ with a clear cross-over around some
intermediate time. Fig.(3) exhibits the asymptotic constancy of
$\sigma_{VV}^2 (t)$ as a function of time for different temperatures. The
effect of correlation time $\tau_c$ on the variance $\sigma_{XX}^2 (t)$ 
has been examined in Fig.(4) for a fixed
high temperature $k_BT$ = 10.0. It is interesting to note that the cross-over 
region gets longer for larger correlation time.

Figs.(5) and (6) illustrate the zero temperature situation. In this regime
non-Markovian effects are strong which is evident from vacuum fluctuations 
growing in time in an oscillatory fashion at early stages
for different values of
correlation time as shown in Fig.(5). In Fig.(6) we show how the initial
growth of variance $\sigma_{VV}^2 (t)$ finally settles down to a 
constant non-thermal energy value.

%%%%%%%%%%%%%%%%%%%%%%%%% SECTION V %%%%%%%%%%%%%%%%%%%%%%%%%%%%%%%%%%%%%

\section{The generalized quantum Fokker-Planck equation}

We now return to our general analysis as carried out in Sec.~{III}.
To write down the Fokker-Planck description for the evolution of probability
density function of quantum mechanical mean values of co-ordinate and 
momentum of the particle it is necessary 
to consider the statistical distribution of 
noise which we assume here to be Gaussian. For Gaussian noise processes we
define the joint characteristic function in terms of the standard mean values
and variances as follows ;
\begin{equation}
\label{eq42}
\tilde{P} ( \mu, \rho, t ) = \exp \left [ i \mu \langle X (t) \rangle_S +
i \rho \langle V (t) \rangle_S - \frac{1}{2} \left \{
\sigma_{XX}^2 (t) \mu^2 + 2 \sigma_{XV}^2 (t) \mu \rho +
\sigma_{VV}^2 (t) \rho^2 \right \} \right ] \; \; .
\end{equation}

\noindent
Using the standard procedure \cite{mazo,adelman} we write down below the
Fokker-Planck equation (FPE) obeyed by the joint probability density function
$P (X,V,t)$ which is the inverse Fourier transform of the characteristic
function :
\begin{eqnarray}
\label{eq43}
\left (
\frac{\partial}{\partial t} + V \frac{\partial}{\partial X} \right )
P (X,V,t) & = & \xi (t) \frac{\partial}{\partial V} V P (X,V,t) + 
\varphi (t) \frac{\partial^2}{\partial V^2} P(X,V,t) \nonumber \\
& & +
\psi (t) \frac{\partial^2}{\partial X \partial V} P(X,V,t)
\end{eqnarray}

\noindent
where
\begin{mathletters}
\begin{eqnarray}
\xi (t) & = & - \dot{h} (t)/ h (t) \; \; , \label{eq44a} \\
\varphi (t) & = & \xi (t) \sigma_{VV}^2 (t) + 
\frac{1}{2} \dot{\sigma}_{VV}^2 (t) \; \; {\rm and} \label{eq44b} \\
\psi (t) & = & - \sigma_{VV}^2 (t) + \xi (t) \sigma_{XV}^2 (t)
+ \dot{\sigma}_{XV}^2 (t) \; \; . \label{eq44c}
\end{eqnarray}
\end{mathletters}

\noindent
The above FPE is the exact quantum mechanical version of the classical 
non-Markovian FPE 
and is valid at any arbitrary temperature and friction.

The decisive advantage of the present approach is again noteworthy. We have
mapped the operator generalized Langevin equation into a generalized Langevin
equation in $c$-numbers (\ref{eq15}) and its equivalent Fokker-Planck 
equation (\ref{eq43}). 
The present approach bypasses the earlier methods of
quasi-probabilistic distribution functions employed widely in quantum optics
over the decades \cite{whl,wlamb,lax,optics,dsr}
in a number of ways. First, unlike the quasi-probabilistic
distribution functions, the probability distribution function $P(X,V,t)$
is valid for non-Markov processes. Second, while the corresponding
characteristic functions for quasi-probabilistic distribution functions are
operators, we make use of characteristic functions which are numbers.
Third, as pointed out earlier the quasi-distribution functions often become 
negative or singular in the strong quantum domain and pose serious 
problems. The present approach is free from such shortcomings since the 
probability density function, $P(X,V,t)$ behaves here as a true probability 
function rather than a quasi-probability function.

%%%%%%%%%%%%%%%%%%%%%%%%%%% SECTION VI %%%%%%%%%%%%%%%%%%%%%%%%%%%%%%%%%%

\section{Generalized Quantum diffusion equation}

In their landmark paper on classical Brownian motion 
Ornstein and Uhlenbeck \cite{geu}
solved the classical Markovian FPE to find $P(X,V,t)$ and then in a bid to
obtain Einstein's diffusion equation tried to 
evaluate $p(X,t)$, the probability density function in configuration space
by integrating over $V$. It was shown that it is difficult, if not impossible
to obtain a differential equation for $P(X,V_0,t)$ from the classical
Markovian FPE which for $t\gg 1/\gamma_0$ would become a diffusion equation.
However, for the classical non-Markovian case Mazo \cite{mazo} in late
seventies addressed this problem
by considering an initial Maxwellian distribution $\Phi (V_0)$
of the initial velocity
$V_0$ and then derived the exact differential equation satisfied by
$p(X,t)$ where
\begin{eqnarray*}
p(X,t) = \int P(X,V_0,t) \; \Phi (V_0) \; dV_0 \; \; .
\end{eqnarray*}

\noindent
The resulting equation thus reduces to the diffusion equation for 
$t \gg 1/\gamma$. We follow Mazo's procedure to derive an exact quantum
mechanical version of the classical non-Markovian case, a differential
equation which for $t \gg 1/\gamma$ goes over into a quantum diffusion 
equation.
To this end we proceed as follows;
from Eq.(\ref{eq42}) for $\rho =0$ case we have
\begin{eqnarray}
\tilde{p} (\mu, t) & = & \int \tilde{p} (\mu, t) \; \Phi (V_0) \; dV_0
\nonumber \\
& = & \exp \left ( -\frac{1}{2} \mu^2 \sigma_{XX}^2 (t) \right ) \;
\exp ( i \mu  X_0 ) \;  \int 
\exp [ i \mu V_0 H (t) ] \; \Phi (V_0) \; dV_0 \; \; . \label{eq45}
\end{eqnarray}

\noindent
Here we take the initial Gaussian distribution of the quantum mechanical mean 
values of the velocity of the particle,
\begin{equation}
\label{eq46}
\Phi (V_0) = \left ( \frac{1}{2\pi \Delta_0} \right )^{1/2} 
\exp \left ( - \frac{V_0^2}{2\Delta_0} \right )
\end{equation}

\noindent
where
\begin{equation}
\label{eq47}
\Delta_0 = \varphi (\infty) / \xi(\infty) \; \; .
\end{equation}

\noindent
It is not difficult to note that the above choice is dictated by the
stationary solution of the QFPE (\ref{eq43}), i.e., (\ref{eq46}) satisfies
(\ref{eq43}) at equilibrium.
The explicit time-dependent expressions for $\varphi (t)$ and $\xi (t)$
have been given in (\ref{eq44a}) and (\ref{eq44b}). Inserting Eq.(\ref{eq46})
in (\ref{eq45}) and then performing the inverse Fourier transform 
after integration over $V_0$ we arrive
at the following equation after little a algebra,
\begin{equation}
\label{eq48}
\frac{\partial p(X,t)}{\partial t} = D_{q} (t)
\frac{\partial^2 p(X,t)}{\partial X^2} \; \; .
\end{equation}

\noindent
This is the quantum analogue of Einstein's diffusion equation where
the explicit structure of the time-dependent quantum diffusion coefficient, 
$D_{q} (t)$ is given by,
\begin{equation}
\label{eq49}
D_{q} (t) = \sigma_{XV}^2 (t) + \Delta_0 H (t) h (t) \; \; .
\end{equation}

\noindent
The required variances, the relaxation functions and other related quantities
in Eq.(\ref{eq49}) are given in (\ref{eq31c}), (\ref{eq26}), (\ref{eq22})
and (\ref{eq47}). We now discuss the limiting cases. For classical
Markovian limit the variance $\sigma_{XV}^2 (t)$ gives $k_BT/\gamma_0$ for
$t \gg 1/\gamma_0$ and the second term in $D_q (t)$ vanishes in the long time 
limit, so that we recover Einstein's diffusion coefficient in configuration
space. In the low temperature, however, the quantum effects begin to dominate. 
It is interesting to note that based on Feynman-Vernon path integral
technique \cite{path1,path2}, Hakim and Ambegaokar \cite{vhva} had considered 
explicit quantum
corrections to classical diffusion to examine the differential behavior
of high and low temperature dependence in the dynamics for 
Leggett-Caldeira initial conditions. 
The non-Markovian nature of the dynamics is taken into account by considering
the frequency dependence of the bath with a suitable low frequency cut-off. 
The transient
behavior in the quantum correction to classical diffusion is therefore
only observable on the timescales longer than the inverse cut-off frequency.
The present treatment being exact, equipped to deal with arbitrary
noise correlation at all temperatures and free from divergences does not
require any such cut-off. The quantum diffusion coefficient can be followed 
arbitrarily from transient to the asymptotic regions.
To explore the associated non-Markovian nature of the dynamics in the
present case it is necessary to go over to numerical evaluation of $D_q (t)$.
In Fig.(7) [ compare with Fig.~1 of Ref.~\cite{vhva} ]
we plot the variation of quantum diffusion coefficient $D_q (t)$
for several values of temperatures
as a function of time for the exponential memory kernel considered in
our example in Sec.~{IV}. It is apparent that while the short time 
behaviour is characterized by a sharp increase followed by a maximum, the 
diffusion coefficient settles down to a constant value in the asymptotic
limit. The short time behaviour is dominated by the 
second term in (\ref{eq49}) due to the relaxation functions $H(t)$ and $h(t)$
of which the latter vanishes in the long time limit.
Again the first term in
(\ref{eq49}) offers no contribution to diffusion coefficient from its
classical part in the vacuum limit at $T=0$. The solid curve in Fig.(7)
thus shows the evolution of a non-thermal diffusion coefficient of pure
quantum origin.

%%%%%%%%%%%%%%%%%%%%%%%%%%%%%% SECTION VII %%%%%%%%%%%%%%%%%%%%%%%%%%%%%%%%

\section{Quantum Smoluchowski equation}

We now consider the diffusion of a particle in an external potential
$V(X)$ as described by QGLE (\ref{eqn1}). In the overdamped limit we
drop the inertial term $\ddot{X} (t)$ and
the damping kernel $\gamma (t-t')$ is reduced to $\gamma_0 \; \delta (t-t')$
for vanishing $\tau_c$ in (\ref{eq34}). $\gamma_0$ is the Markovian
limit of dissipation. Eq.(\ref{eqn1}) then assumes the following form
\begin{equation}
\label{eqs1}
\dot{X} + \frac{1}{\gamma_0} [ V' ( X ) - Q ( X , t ) ]
= \frac{F(t)}{\gamma_0} \; \; .
\end{equation}

\noindent
Expressing $V'(X)-Q(X,t)$ as a derivative of an effective quantum potential
$V_{quant} (X,t)$ with respect to $X$, the equivalent description in terms 
of true probability distribution function $p(X,t)$ is given by

\begin{mathletters}

\begin{equation}
\label{eqs2}
\frac{\partial p(X,t) }{\partial t} = \frac{1}{\gamma_0}
\frac{\partial }{\partial X} \left [ V_{quant}'(X,t) p (X,t) \right ]
+ D_{qo} \frac{\partial^2 p}{\partial X^2} \; \; .
\end{equation}

\noindent
with
\begin{equation}
V_{quant}' (X,t) = V'(X) - Q(X,t)
\end{equation}

\end{mathletters}

\noindent
where $Q(X,t)$ is defined in (13).
Here $D_{qo}$ is the quantum diffusion coefficient in the overdamped limit
which can be obtained with the help of the following definition \cite{whl}
\begin{equation}
\label{eqs3}
2 D_{qo} = \frac{1}{\Delta t} \int_t^{t+\Delta t} dt_1
\int_t^{t+\Delta t} dt_2 \; \frac{1}{\gamma_0^2}
\langle F (t_1) F (t_2) \rangle_S \; \; .
\end{equation}

\noindent
Here the correlation function
$\langle F (t_1 ) F (t_2) \rangle_S / \gamma_0^2 $ of the $c$-number
quantum noise is given by
Eq.(\ref{eq17}) in the continuum limit. We then
make use of
Eq.(\ref{eq33}) for vanishing $\tau_c$ in (\ref{eqs3}) to obtain
after explicit integration
\begin{equation}
\label{eqs4}
D_{qo} = \frac{1}{2\gamma_0} \hbar \tilde{\omega}
[ 2 \bar{n} ( \tilde{\omega} ) + 1 ]
\end{equation}

\noindent
where the frequency $\tilde{\omega}$ in (\ref{eqs4}) refers to linearized
frequency of the nonlinear system \cite{whl}. 
We now discuss the classical and vacuum limits of the quantum Smoluchowski
equation (\ref{eqs2}). It is easy to check that in the limit
$\hbar \tilde{\omega} \ll k_BT$, $D_{qo}$ reduces to Einstein's classical
diffusion coefficient $k_BT/\gamma_0$. At the same time $Q(X,t)$ vanishes 
so that $V_{quant}' (X,t)$ goes over to $V'(X)$ and one recovers the usual 
classical Smoluchowski equation. In the opposite limit as 
$T \rightarrow 0$, however, both quantum noise due to nonlinearity of the
system and vacuum fluctuation orginating from the heat bath make significant
contribution. $D_{qo}$ in this limit assumes the form 
$\hbar \tilde{\omega}/2\gamma_0$. In this context we refer to a recent
treatment on large friction limit in quantum dissipative dynamics \cite{apg}
to point out that the latter theory does not retain its full validity as
$T \rightarrow 0$ since the quantum noise of the heat bath disappears in the
vacuum limit.

The second noteworthy feature about the quantum Smoluchowski equation
(\ref{eqs2}) is that unlike Wigner function based equations \cite{whz}
it does not contain higher order (higher than second) derivatives of 
$p(X,t)$. The positive definiteness of the probability distribution
function is thus ensured.

It is important to emphasize at this juncture that so far as the general
formulation of the theory is concerned, Eq.(\ref{eqs2}) contains quantum
corrections to all
orders. In this sense Eq.(\ref{eqs2}) is formally an exact quantum
analogue of classical Smoluchowski equation. To make it more explicit we 
return to the quantum mechanics of the system in Heisenberg picture
to write the operators $\hat{x}$ and $\hat{p}$ as
\begin{equation}
\label{eqn3}
\hat{x} (t) = \langle \hat{x} (t) \rangle + \delta \hat{x} \; \;
{\rm and} \; \; 
\hat{p} (t) = \langle \hat{p} (t) \rangle + \delta \hat{p} \; \; .
\end{equation}

\noindent
$\langle \hat{x} (t) \rangle$ and $\langle \hat{p} (t) \rangle$ are the
quantities signifying quantum mechanical averages and $\delta \hat{x}$
and $\delta \hat{p}$ are quantum corrections. By construction
$\langle \delta \hat{x} \rangle$ and $\langle \delta \hat{p} \rangle$
are zero and they obey the commutation
relation [$\delta \hat{x}, \delta \hat{p}$] = $i\hbar$. Using (\ref{eqn3})
in $\langle V' (\hat{x}) \rangle$ and a Taylor expansion around
$\langle \hat{x} \rangle$ it is possible to express $Q(X,t)$ as 
[ see Eq.(13) ]:
\begin{mathletters}
\begin{equation}
\label{eqn4}
Q (X,t) = - \sum_{n \geq 2} \frac{1}{n!} V_{n+1} (X)
\langle \delta \hat{x}^n (t) \rangle
\end{equation}

\noindent
where $V_n (X)$ is the $n$-th derivative of the potential at
$X ( \equiv \langle \hat{x} \rangle $). Eq.(\ref{eqn4}) suggests a simple
expression for an effective potential $V_{quant} (X, t)$ as
\begin{equation}
\label{eqn4a}
V_{quant} (X,t) = V (X) + \sum_{n \geq 2} \frac{1}{n!} V_n(X)
\langle \delta \hat{x}^n (t) \rangle
\end{equation}
\end{mathletters}

\noindent
where the classical potential $V(X)$ gets modified by the quantum
corrections {\it to all orders}. To solve quantum Smoluchowski equation it is
therefore necessary to calculate
$\langle \delta \hat{x}^2 (t) \rangle$,
$\langle \delta \hat{x}^3 (t) \rangle$, etc. To the lowest order
$\langle \hat{x} \rangle$ and $\langle \delta \hat{x}^2 \rangle$
follow a coupled set of equations as given below
\begin{mathletters}
\begin{eqnarray}
\label{eqn5a}
\frac{d}{dt} \langle \hat{x} \rangle & = & \langle \hat{p} \rangle 
\\
\label{eqn5b}
\frac{d}{dt} \langle \hat{p} \rangle & = & -V' ( \langle \hat{x} \rangle )
- \frac{1}{2} V''' ( \langle \hat{x} \rangle )
\langle \delta \hat{x}^2 \rangle
\\
\label{eqn5c}
\frac{d}{dt} \langle \delta \hat{x}^2 \rangle & = & 
\langle \delta \hat{x}
\delta \hat{p} + \delta \hat{p} \delta \hat{x} \rangle 
\\
\label{eqn5d}
\frac{d}{dt} \langle \delta \hat{x}
\delta \hat{p} + \delta \hat{p} \delta \hat{x} \rangle & = &
2 \langle \delta \hat{p}^2 \rangle - 2 V'' ( \langle \hat{x} \rangle )
\langle \delta \hat{x}^2 \rangle
\\
\label{eqn5e}
\frac{d}{dt} \langle \delta \hat{p}^2 \rangle & = & 
- V'' ( \langle \hat{x} \rangle ) \langle \delta \hat{x}
\delta \hat{p} + \delta \hat{p} \delta \hat{x} \rangle  \; \; .
\end{eqnarray}
\end{mathletters}

\noindent
The above set of equations can be derived \cite{sm}
from the Heisenberg's equation
of motion. If one is interested in the local dynamics around a point (say,
at the bottom or top of the potential well), 
the set of equations get decoupled and it is easy to obtain simple
analytic solutions of (\ref{eqn5a})-(\ref{eqn5e}) for
$\langle \hat{x} \rangle$ and $\langle \delta \hat{x}^2 \rangle$
for (\ref{eqn4}). The higher order estimates (e.g., fourth order)
of the quantum corrections can be obtained from the solutions of the
equations of successive higher order derived earlier by Sundaram and
Milonni \cite{sm} or otherwise \cite{akp}. 
Since the quantum corrections due to the system are
calculated by different sets of equations for succesive orders,
the measure of accuracy of truncation can be understood easily. It is,
therefore, obvious that the present scheme provides a simple, systematic
and quantitative estimate of the mean field and other decorrelation methods
on the basis of quantum-classical correspondence.

%%%%%%%%%%%%%%%%%%%%%%%%%%%%%% SECTION VIII %%%%%%%%%%%%%%%%%%%%%%%%%%%%%%%%

\section{Conclusions}

The main purpose of this paper is to enquire whether a stochastic differential
equation in $c$-numbers in the form of a generalized Langevin equation and
its corresponding Fokker-Planck equation and diffusion equation
and Smoluchowski equation in terms of
{\it true probability functions} are viable for description of
non-Markovian quantum Brownian motion. Based on an initial coherent state
representation of bath oscillators and an equilibrium distribution of
quantum mechanical mean values of their co-ordinates and momenta, 
which satisfy the
essential properties of the associated noise of the bath degrees of
freedom, we derive a QGLE for quantum mechanical mean value of the position
of the particle. The main conclusions of this study are the 
following :

(i) Our QGLE (\ref{eq15}) is amenable to analysis in terms of the 
methods developed earlier for the treatment of classical non-Markovian
theory of Brownian motion.

(ii) The generalized Langevin equation (\ref{eqn1}), the corresponding  
Fokker-Planck equation (\ref{eq43}) and the diffusion equation 
(\ref{eq48}) and also the Smoluchowski equation (\ref{eqs2})
are the {\it exact quantum analogues} of their classical
versions \cite{mazo,adelman}. The probability distribution functions 
as employed here
bear the {\it true notion of statistical probability} rather than 
that of quasi-probability.

(iii) The theory of quantum Brownian motion developed here is valid for
arbitrary {\it noise correlation} and {\it temperature} and is free from
divergences.

(iv) The realization of noise as a classical-looking entity which satisfies
quantum fluctuation-dissipation relationship (\ref{eq12}) allows
ourselves to envisage quantum Brownian motion as a natural extension of its
classical conterpart. The method is based on canonical quantization procedure 
and makes no reference to path integral formulations.

We conclude by mentioning that the method discussed here is promising for 
simple differential equation based approaches \cite{jrc} to quantum activated
processes, tunneling problems as shown elsewhere \cite{dbsd}, quantum
ratchet \cite{qrat1,qrat2,qrat3} and in problems relating to the motion 
in periodic fields \cite{rat1,rat2,rat3,rat4} and allied issues.

\acknowledgments
The authors are indebted to the Council of Scientific and Industrial Research
(C.S.I.R.), Government of India for financial support.

%%%%%%%%%%%%%%%%%%%%%%%%%%%%%% APPENDIX %%%%%%%%%%%%%%%%%%%%%%%%%%%%%%%%%%%

\begin{appendix}

\section{The explicit forms of  ${\cal F}_X (\omega, t)$ and
${\cal F}_V (\omega, t)$ }

${\cal F}_X (\omega, t)$ consists of eleven terms which are given below ;

\begin{eqnarray}
{\cal F}_X (\omega, t) & = & {\cal F}^{(1)}_X (\omega, t) +
{\cal F}^{(2)}_X (\omega, t) + {\cal F}^{(3)}_X (\omega, t) +
{\cal F}^{(4)}_X (\omega, t) + {\cal F}^{(5)}_X (\omega, t) \nonumber \\
& &  + {\cal F}^{(6)}_X (\omega, t)
+ {\cal F}^{(7)}_X (\omega, t) + {\cal F}^{(8)}_X (\omega, t) 
+ {\cal F}^{(9)}_X (\omega, t) + {\cal F}^{(10)}_X (\omega, t) \nonumber \\
& & + {\cal F}^{(11)}_X (\omega, t) \; \; .
\end{eqnarray}

\noindent
The explicit structures of ${\cal F}^{(i)}_X (\omega, t)$  
($i=1,\ldots,11)$ are given by

\begin{equation}
{\cal F}^{(1)}_X (\omega, t) = 
\frac{1}{\gamma_0 \omega^2} (1-\cos \omega t) \; \; ,
\end{equation}

\begin{eqnarray}
{\cal F}^{(2)}_X (\omega, t) & = & 
\frac{{\cal A} A_3^{(\omega)} }{\gamma_0 \omega} [ \cos (\alpha + \omega t) - \cos \alpha ]
- \frac{{\cal A} A_4^{(\omega)} }{\gamma_0 \omega} [ \cos (\alpha - \omega t) - \cos \alpha ]
\nonumber \\
& &
- \frac{{\cal A} A_5^{(\omega)} }{\gamma_0 \omega} [ \sin (\alpha + \omega t) - \sin \alpha ]
+ \frac{{\cal A} A_6^{(\omega)} }{\gamma_0 \omega} [ \sin (\alpha - \omega t) - \sin \alpha ] 
\; \; ,
\end{eqnarray}

\begin{eqnarray}
{\cal F}^{(3)}_X (\omega, t) & = & 
-\frac{{\cal A} A_1^{(\omega)} }{2 \gamma_0^2}
[ e^{-t/2 \tau_c} \left \{ \sin ( \lambda t + \alpha ) +
2 \lambda \tau_c \cos ( \lambda t + \alpha ) \right \}  \nonumber \\
& & - \left \{ \sin \alpha + 
2 \lambda \tau_c \cos \alpha \right \} ] \; \; ,
\end{eqnarray}

\begin{eqnarray}
{\cal F}^{(4)}_X (\omega, t) & = & 
-\frac{{\cal A} A_2^{(\omega)} }{2 \gamma_0^2}
[ e^{-t/2 \tau_c} \left \{ \cos ( \lambda t + \alpha ) -
2 \lambda \tau_c \sin ( \lambda t + \alpha ) \right \}  \nonumber \\
& & - \left \{ \cos \alpha 
- 2 \lambda \tau_c \sin \alpha \right \} ] \; \; ,
\end{eqnarray}

\begin{eqnarray}
{\cal F}^{(5)}_X (\omega, t) & = & \frac{{\cal A}^2 A_2^{(\omega)} }{8 \gamma_0^2}
\left [ e^{-t/\tau_c} \left \{ \sin 2(\lambda t + \alpha) + 2 \lambda \tau_c
\cos 2(\lambda t + \alpha) \right \} \right. \nonumber \\
& & \left. - \left \{ \sin 2 \alpha + 2 \lambda \tau_c
\cos 2 \alpha \right \} \right ] \; \; ,
\end{eqnarray}

\begin{eqnarray}
{\cal F}^{(6)}_X (\omega, t) & = & {\cal A}^2 A_1^{(\omega)} 
\left ( \frac{\tau_c}{2 \gamma_0} \right ) 
\left [ e^{-t/\tau_c} + \frac{ e^{-t/\tau_c} }{4 \gamma_0 \tau_c}
\left \{ 2 \lambda \tau_c \sin 2 (\lambda t + \alpha) - 
\cos 2 (\lambda t + \alpha ) \right \} \right. \nonumber \\
& & \left. - \left \{ 1 + \frac{1}{4 \gamma_0 \tau_c} \left (
2 \lambda \tau_c \sin 2\alpha - \cos 2\alpha \right ) \right \} \right ]
\; \; ,
\end{eqnarray}

\begin{eqnarray}
{\cal F}^{(7)}_X (\omega, t) & = & -\frac{ {\cal A} }{\gamma_0 \omega}
[  A_3^{(\omega)} \{ e^{-t/2\tau_c} ( 2 \tau_c (\lambda-\omega)
\sin [ \alpha + ( \lambda-\omega ) t ] - 
\cos [ \alpha + ( \lambda-\omega ) t] )  \nonumber \\
& & - 2 \tau_c ( \lambda - \omega ) \sin \alpha + \cos \alpha \}
\nonumber \\
& & - A_4^{(\omega)} \{ e^{-t/2\tau_c} ( 
2 \tau_c (\lambda+\omega)
\sin [ \alpha + ( \lambda+\omega ) t ] - 
\cos [ \alpha + ( \lambda+\omega ) t] ) \nonumber \\
& & - 2 \tau_c ( \lambda + \omega ) \sin \alpha + \cos \alpha \} ]
\; \; ,
\end{eqnarray}

\begin{eqnarray}
{\cal F}^{(8)}_X (\omega, t) & = & 
\frac{ {\cal A}^2 A_3^{(\omega)} }{\gamma_0} [ A_3^{(\omega)} \{
e^{-t/2\tau_c} ( 2 \tau_c (\lambda-\omega) \sin (\lambda-\omega) t -
\cos (\lambda-\omega )t ) + 1 \} \nonumber \\
& & - A_4^{(\omega)} \{ e^{-t/2\tau_c} ( 2 \tau_c (\lambda +\omega)
\sin [ 2\alpha + (\lambda+\omega) t ] - 
\cos [ 2\alpha + (\lambda+\omega) t ] ) \nonumber \\
& & - ( 2 \tau_c (\lambda+\omega) \sin 2 \alpha - \cos 2\alpha ) \} ] \; \; ,
\end{eqnarray}

\begin{eqnarray}
{\cal F}^{(9)}_X (\omega, t) & = & 
\frac{ {\cal A}^2 A_4^{(\omega)} }{\gamma_0} [ A_4^{(\omega)} \{
e^{-t/2\tau_c} ( 2 \tau_c (\lambda+\omega) \sin (\lambda+\omega) t -
\cos (\lambda+\omega )t ) + 1 \} \nonumber \\
& & - A_3^{(\omega)} \{ e^{-t/2\tau_c} ( 2 \tau_c (\lambda -\omega)
\sin [ 2\alpha + (\lambda-\omega) t ] - 
\cos [ 2\alpha + (\lambda-\omega) t ] ) \nonumber \\
& & - ( 2 \tau_c (\lambda-\omega) \sin 2 \alpha - \cos 2\alpha ) \} ] \; \; ,
\end{eqnarray}

\begin{eqnarray}
{\cal F}^{(10)}_X (\omega, t) & = & -
\frac{ {\cal A}^2 A_5^{(\omega)} }{\gamma_0} [ A_4^{(\omega)} \{
e^{-t/2\tau_c} ( \sin [ 2 \alpha + (\lambda+\omega) t ] +
2\tau_c (\lambda+\omega) \cos [ 2\alpha + (\lambda + \omega) t] )
\nonumber \\
& & - ( \sin 2 \alpha + 2\tau_c (\lambda+\omega) \cos 2 \alpha ) \}
\nonumber \\
& & + A_3^{(\omega)} \{ e^{-t/2\tau_c} ( \sin (\lambda-\omega)t +
2\tau_c (\lambda-\omega) \cos (\lambda-\omega)t ) - 
2 \tau_c (\lambda-\omega ) \} ] 
\end{eqnarray}

\noindent
and

\begin{eqnarray}
{\cal F}^{(11)}_X (\omega, t) & = & -
\frac{ {\cal A}^2 A_6^{(\omega)} }{\gamma_0} [ A_3^{(\omega)} \{
e^{-t/2\tau_c} ( \sin [ 2 \alpha + (\lambda-\omega) t ] +
2\tau_c (\lambda-\omega) \cos [ 2\alpha + (\lambda - \omega) t] )
\nonumber \\
& & - ( \sin 2 \alpha + 2\tau_c (\lambda-\omega) \cos 2 \alpha ) \}
\nonumber \\
& & + A_4^{(\omega)} \{ e^{-t/2\tau_c} ( \sin (\lambda+\omega)t +
2\tau_c (\lambda+\omega) \cos (\lambda+\omega)t ) - 
2 \tau_c (\lambda+\omega ) \} ] 
\end{eqnarray}

\noindent
where
\begin{eqnarray}
A_1^{(\omega)} = \tau_c \left [ 
\frac{1}{ 1 + 4 \tau_c^2 (\lambda-\omega)^2} +
\frac{1}{ 1 + 4 \tau_c^2 (\lambda+\omega)^2} \right ] \; \; ,
\nonumber \\
A_2^{(\omega)} = 2\tau_c^2 \left [ 
\frac{\lambda-\omega}{ 1 + 4 \tau_c^2 (\lambda-\omega)^2} +
\frac{\lambda-\omega}{ 1 + 4 \tau_c^2 (\lambda+\omega)^2} \right ] \; \; ,
\nonumber\\
A_3^{(\omega)} = \frac{\tau_c}{1 + 4 \tau_c^2 (\lambda-\omega)^2}
\; \; , \; \;
A_4^{(\omega)} = \frac{\tau_c}{1 + 4 \tau_c^2 (\lambda+\omega)^2}
\; \; , 
\nonumber \\
A_5^{(\omega)} = 
\frac{2 \tau_c^2 (\lambda-\omega)}{1 + 4 \tau_c^2 (\lambda-\omega)^2}
\; \; {\rm and} \; \;
A_6^{(\omega)} = 
\frac{2 \tau_c^2 (\lambda+\omega)}{1 + 4 \tau_c^2 (\lambda+\omega)^2}
\; \; .
\end{eqnarray}

Similarly we have
\begin{eqnarray}
{\cal F}_V (\omega, t) & = & {\cal F}^{(1)}_V (\omega, t) +
{\cal F}^{(2)}_V (\omega, t) + {\cal F}^{(3)}_V (\omega, t) +
{\cal F}^{(4)}_V (\omega, t) + {\cal F}^{(5)}_V (\omega, t) \nonumber \\
& & + {\cal F}^{(6)}_V (\omega, t)  + {\cal F}^{(7)}_V (\omega, t) 
\end{eqnarray}

\noindent
with
\begin{eqnarray}
{\cal F}^{(1)}_V (\omega, t) & = & \frac{1}{4} \left (
\frac{ A_1^{(\omega)} }{2\tau_c} + \lambda A_2^{(\omega)} \right )
\left [ e^{-t/\tau_c} + \frac{ e^{-t/\tau_c} }{4\gamma_0\tau_c}
\left \{ 2 \lambda \tau_c \sin 2 (\lambda t + \alpha) - \cos 2
(\lambda t + \alpha ) \right \} \right. \nonumber \\
& & \left. - \left \{ 1 + \frac{1}{4\gamma_0\tau_c} \left (
2 \lambda \tau_c \sin 2 \alpha - \cos 2 \alpha \right ) \right \}
\right ] \; \; ,
\end{eqnarray}

\begin{eqnarray}
{\cal F}^{(2)}_V (\omega, t) & = & \frac{\lambda \tau_c}{2} 
\left (
\lambda A_1^{(\omega)} - \frac{A_2^{(\omega)} }{2\tau_c} \right )
\left [ e^{-t/\tau_c} - \frac{ e^{-t/\tau_c} }{4\gamma_0\tau_c}
\left \{ 2 \lambda \tau_c \sin 2 (\lambda t + \alpha) - \cos 2
(\lambda t + \alpha ) \right \} \right. \nonumber \\
& & \left. - \left \{ 1 - \frac{1}{4\gamma_0\tau_c} \left (
2 \lambda \tau_c \sin 2 \alpha - \cos 2 \alpha \right ) \right \}
\right ] \; \; ,
\end{eqnarray}

\begin{eqnarray}
{\cal F}^{(3)}_V (\omega, t) & = & \frac{-1}{8\gamma_0}
\left ( \frac{ \lambda A_1^{(\omega)} }{\tau_c} +
\lambda^2 A_2^{(\omega)} - \frac{A_2^{(\omega)} }{4\tau_c^2} \right )
\nonumber \\
& & \times \; \left [ e^{-t/\tau_c} \left \{ \sin 2 (\lambda t + \alpha )
+ 2 \lambda \tau_c \cos 2 ( \lambda t + \alpha ) \right \}
- \left \{ \sin 2 \alpha + 2 \lambda \tau_c \cos 2 \alpha \right \}
\right ] \; \; ,
\end{eqnarray}

\begin{eqnarray}
{\cal F}^{(4)}_V (\omega, t) & = & \left (
\frac{ A_3^{(\omega)} }{2\tau_c} + \lambda A_5^{(\omega)} \right )
\left [ \frac{ A_3^{(\omega)} }{2\tau_c} \left \{ e^{-t/2\tau_c} \left (
2 \tau_c (\lambda-\omega) \sin (\lambda-\omega) t - \cos (\lambda-\omega) t
\right ) + 1 \right \} \right. \nonumber \\
& & + \lambda A_3^{(\omega)} \{ e^{-t/2\tau_c} ( \sin
[ 2\alpha + (\lambda-\omega) t] + 2\tau_c (\lambda-\omega) \cos
[ 2\alpha + (\lambda-\omega) t] ) \nonumber \\
& & - e^{-t/2\tau_c} ( \sin (\lambda-\omega)t + 2\tau_c (\lambda-\omega)
\cos (\lambda-\omega)t ) \nonumber \\
& & - ( \sin 2 \alpha + 2\tau_c (\lambda-\omega) \cos 2 \alpha ) +
2 \tau_c (\lambda-\omega) \} \nonumber \\
& & - \frac{ A_4^{(\omega)} }{2\tau_c} \{ e^{-t/2\tau_c} ( 
2 \tau_c (\lambda+\omega) \sin [ 2\alpha + (\lambda+\omega)t] - \cos
[2\alpha + (\lambda+\omega)t ] ) \nonumber \\
& & - ( 2 \tau_c (\lambda+\omega) \sin 2 \alpha - \cos 2 \alpha ) \} ]
\; \; ,
\end{eqnarray}

\begin{eqnarray}
{\cal F}^{(5)}_V (\omega, t) & = & \left (
\frac{ A_4^{(\omega)} }{2\tau_c} + \lambda A_6^{(\omega)} \right )
\left [ \frac{ A_4^{(\omega)} }{2\tau_c} \{ e^{-t/2\tau_c} ( 2\tau_c
(\lambda+\omega) \sin (\lambda+\omega) t - \cos (\lambda+\omega) t ) + 1
\} \right. \nonumber \\
& & - \frac{ A_3^{(\omega)} }{2\tau_c} \{ e^{-t/2\tau_c} ( 2 \tau_c 
(\lambda - \omega) \sin [ 2\alpha + (\lambda - \omega) t ] - \cos
[ 2 \alpha + (\lambda - \omega) t ] ) \nonumber \\
& & - ( 2 \tau_c (\lambda - \omega) \sin 2 \alpha - \cos 2 \alpha ) \}
\nonumber \\
& & + \lambda A_3^{(\omega)} \{ e^{-t/2\tau_c} ( \sin 
[ 2\alpha + (\lambda - \omega) t ] + 2 \tau_c (\lambda - \omega) \cos
[ 2\alpha + (\lambda - \omega)t] ) \nonumber \\
& & - ( \sin 2 \alpha + 2 \tau_c (\lambda - \omega) \cos 2 \alpha ) \}
\nonumber \\
& & - \lambda A_4^{(\omega)} \{ e^{-t/2\tau_c} ( \sin (\lambda + \omega) t
+ 2 \tau_c (\lambda + \omega) \cos (\lambda + \omega) t ) - 2 \tau_c 
(\lambda + \omega) \} ] \; \; ,
\end{eqnarray}

\begin{eqnarray}
{\cal F}^{(6)}_V (\omega, t) & = & \left (
\lambda A_3^{(\omega)}- \frac{ A_5^{(\omega)} }{2\tau_c}  \right )
\left [ \frac{ A_3^{(\omega)} }{2\tau_c} \{ e^{-t/2\tau_c} ( 
\sin (\lambda - \omega) t + 2 \tau_c (\lambda-\omega) 
\cos (\lambda-\omega) t )  \right. \nonumber \\
& & - 2 \tau_c (\lambda-\omega) \} \nonumber \\
& & + \frac{ A_4^{(\omega)} }{2\tau_c} \{ e^{-t/2\tau_c} ( 
\sin [ 2\alpha + (\lambda + \omega) t ] +
2 \tau_c (\lambda + \omega) \cos
[ 2 \alpha + (\lambda + \omega) t ] ) \nonumber \\
& & - ( \sin 2 \alpha + 2 \tau_c (\lambda + \omega) \cos 2 \alpha ) \}
\nonumber \\
& & + \lambda A_4^{(\omega)} \{ e^{-t/2\tau_c} ( 
2 \tau_c (\lambda + \omega) \sin [ 2\alpha + (\lambda + \omega)t] -
\cos [ 2\alpha + (\lambda + \omega) t ] 
) \nonumber \\
& & - ( 2 \tau_c (\lambda + \omega) \sin 2 \alpha - \cos 2 \alpha ) \}
\nonumber \\
& & + \lambda A_3^{(\omega)} \{ e^{-t/2\tau_c} ( 2 \tau_c (\lambda - \omega) 
\sin (\lambda - \omega) t - \cos (\lambda - \omega) t ) + 1  \} ] 
\end{eqnarray}

\noindent
and

\begin{eqnarray}
{\cal F}^{(7)}_V (\omega, t) & = & \left (
\lambda A_4^{(\omega)}- \frac{ A_6^{(\omega)} }{2\tau_c}  \right )
\left [  \frac{ A_3^{(\omega)} }{\tau_c} \{ e^{-t/2\tau_c} ( \sin
[2\alpha + (\lambda-\omega)t]  \right. \nonumber \\
& & + 2 \tau_c (\lambda-\omega) \cos [2\alpha + (\lambda-\omega)t] ) 
- ( \sin 2 \alpha + 2 \tau_c (\lambda-\omega) \cos 2 \alpha ) \}
\nonumber \\
& & + \frac{ A_4^{(\omega)} }{\tau_c} \{ e^{-t/2\tau_c} ( \sin 
(\lambda+\omega) t + 2 \tau_c (\lambda+\omega) \cos (\lambda+\omega) t )
\nonumber \\
& & - 2 \tau_c (\lambda+\omega) \} \nonumber \\
& & + \lambda A_3^{(\omega)} \{ e^{-t/2\tau_c} ( 2 \tau_c (\lambda-\omega)
\sin [ 2\alpha + (\lambda-\omega)t] -\cos [2\alpha + (\lambda-\omega)t] )
\nonumber \\
& & - ( 2 \tau_c (\lambda-\omega) \sin 2 \alpha - \cos 2 \alpha ) \}
\nonumber \\
& & + \lambda A_4^{(\omega)} \{ e^{-t/2\tau_c} ( 2 \tau_c (\lambda+\omega)
\sin (\lambda+\omega) t - \cos (\lambda+\omega) t ) + 1 \} ] \; \; .
\end{eqnarray}

\end{appendix}

%%%%%%%%%%%%%%%%%%%%%%%%%%%% REFERENCES %%%%%%%%%%%%%%%%%%%%%%%%%%%%%%%%

\begin{figure}
\caption{
Plot of $\sigma_{XX}^2 (t)$ against time to show the short time 
behaviour of the variances for different temperatures with fixed parameters
$\gamma_0$ = 1.0 and $\tau_c$ = 1.0. [ Inset : The same as in the main
figure but for a higher temperature, $k_BT$ = 10.0 ]
(units are arbitrary).
}
\end{figure}

\begin{figure}
\caption{
Plot of $\sigma_{XX}^2 (t)$ against time to show long time 
behaviour of the variances for different temperatures. Other parameters are
same as in Fig.(1). [ Inset : The same as in the main
figure but for a higher temperature, $k_BT$ = 10.0 ]
(units are arbitrary).
}
\end{figure}

\begin{figure}
\caption{
Plot of $\sigma_{VV}^2 (t)$ against time to show long time 
behaviour of the variances for different temperatures. Other parameters are
same as in Fig.(1)
(units are arbitrary).
}
\end{figure}

\begin{figure}
\caption{
Plot of $\sigma_{XX}^2 (t)$ against time for different 
correlation times, $\tau_c$ with fixed parameters
$\gamma_0$ = 1.0 and $k_BT$ = 10.0
(units are arbitrary).
}
\end{figure}

\begin{figure}
\caption{
Same as in Fig.(4) but for $k_BT$ = 0.0
(units are arbitrary).
}
\end{figure}

\begin{figure}
\caption{
Plot of $\sigma_{VV}^2 (t)$ against time to show long time 
behaviour due to vacuum fluctuations. Other parameters are
same as in Fig.(1)
(units are arbitrary).
}
\end{figure}

\begin{figure}
\caption{
Plot of quantum diffusion coefficient $D_q (t)$ against time for different 
temperatures and for $\gamma_0$ = 0.275 and $\tau_c$ = 1.0.
[ Inset : Same as in the main figure but for a higher temperature
$k_BT$ = 10.0 ]
(units are arbitrary).
}
\end{figure}

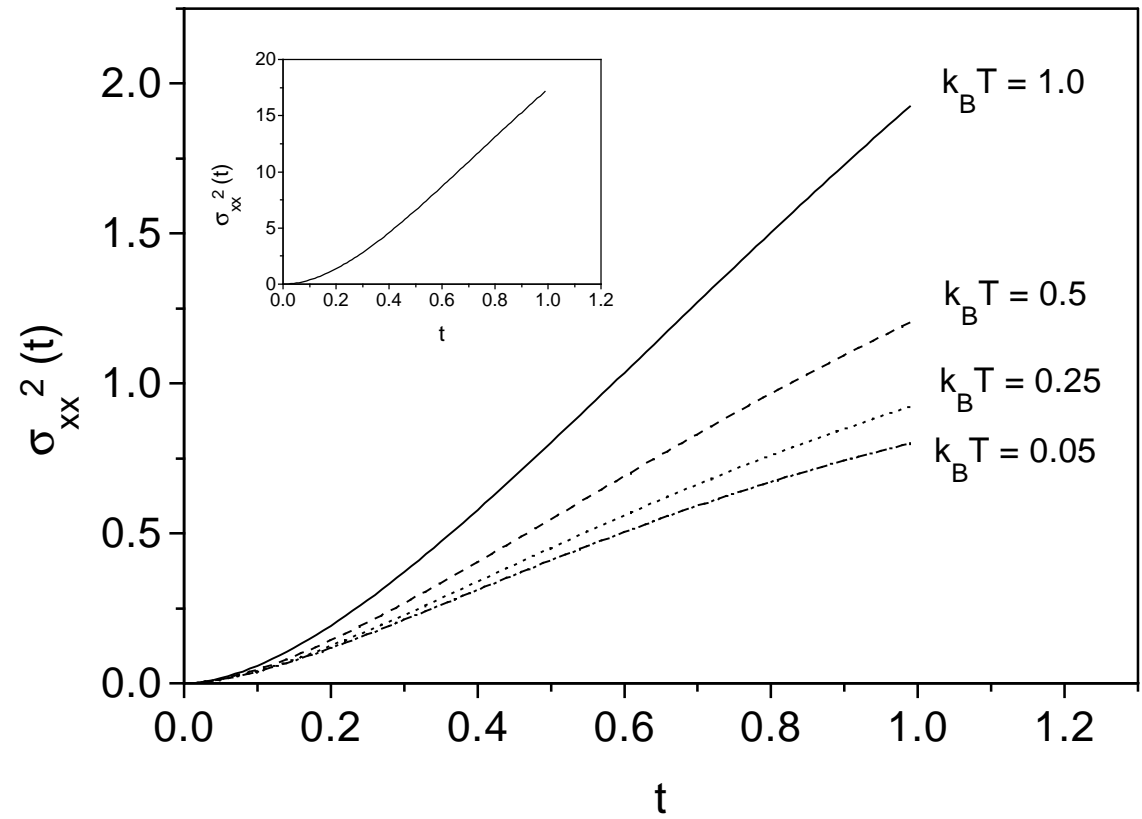
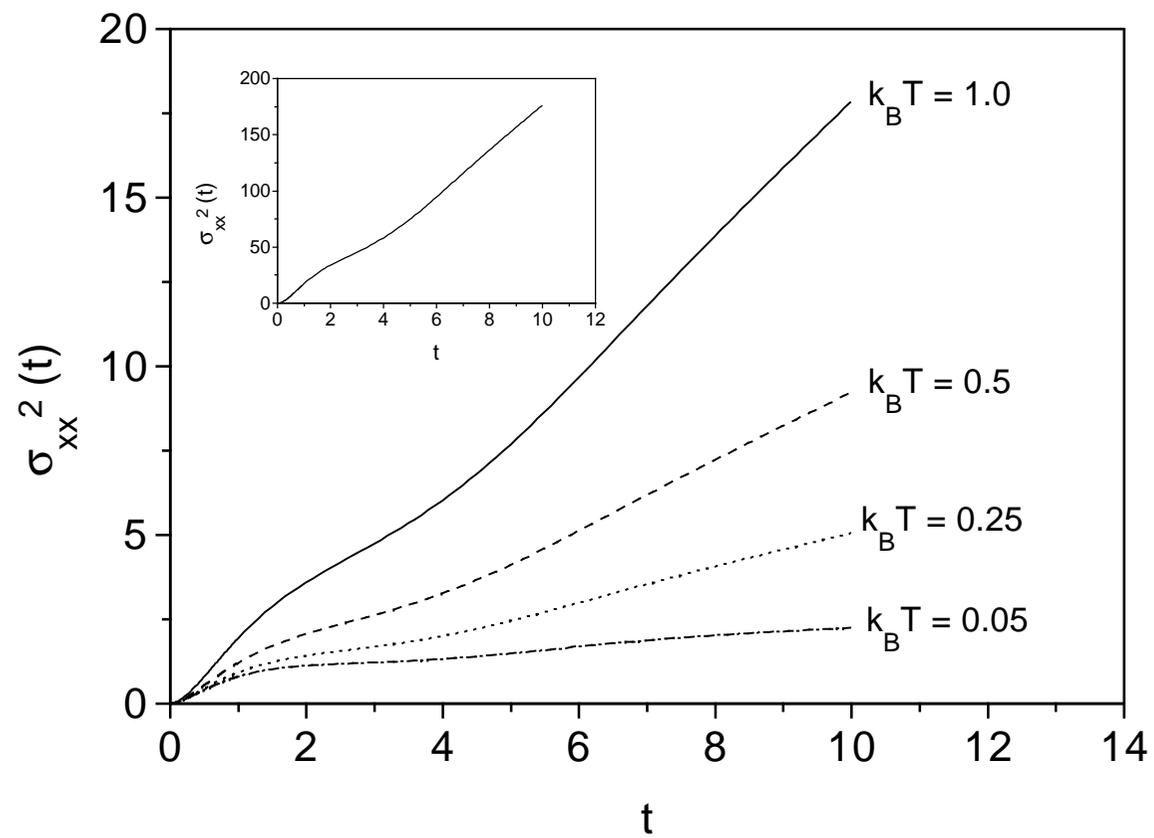
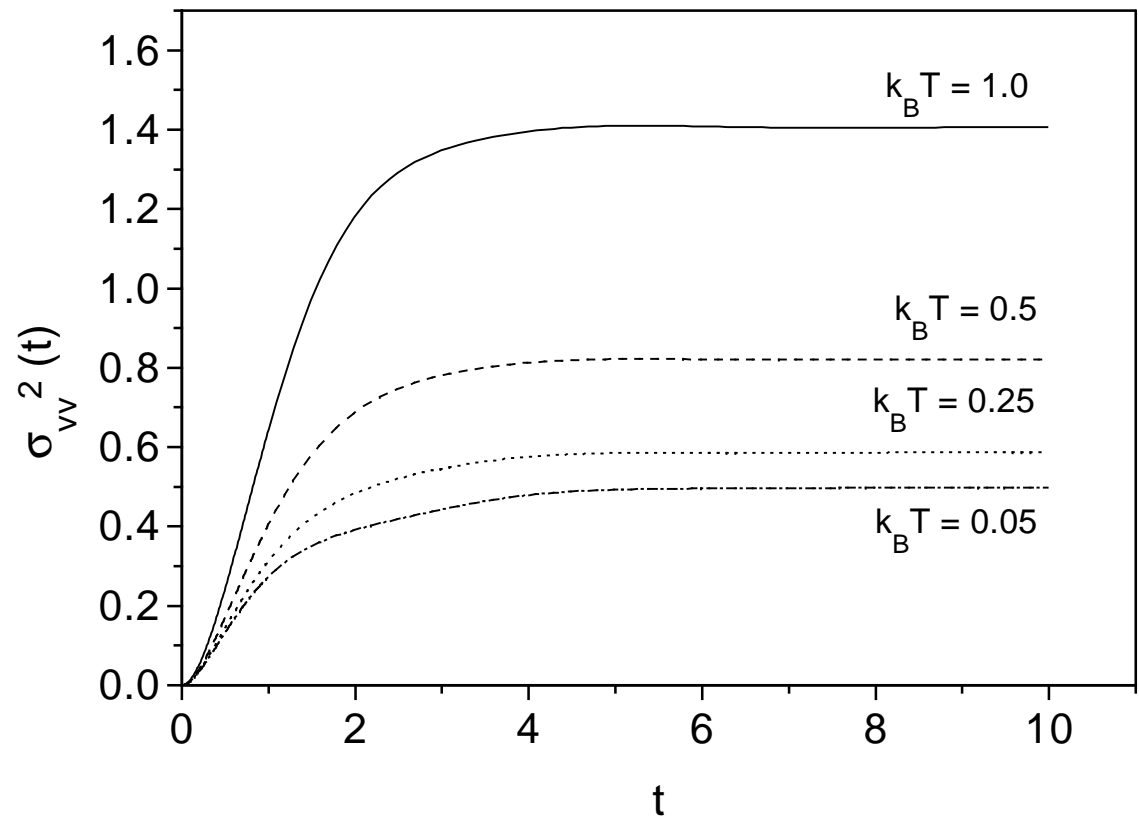
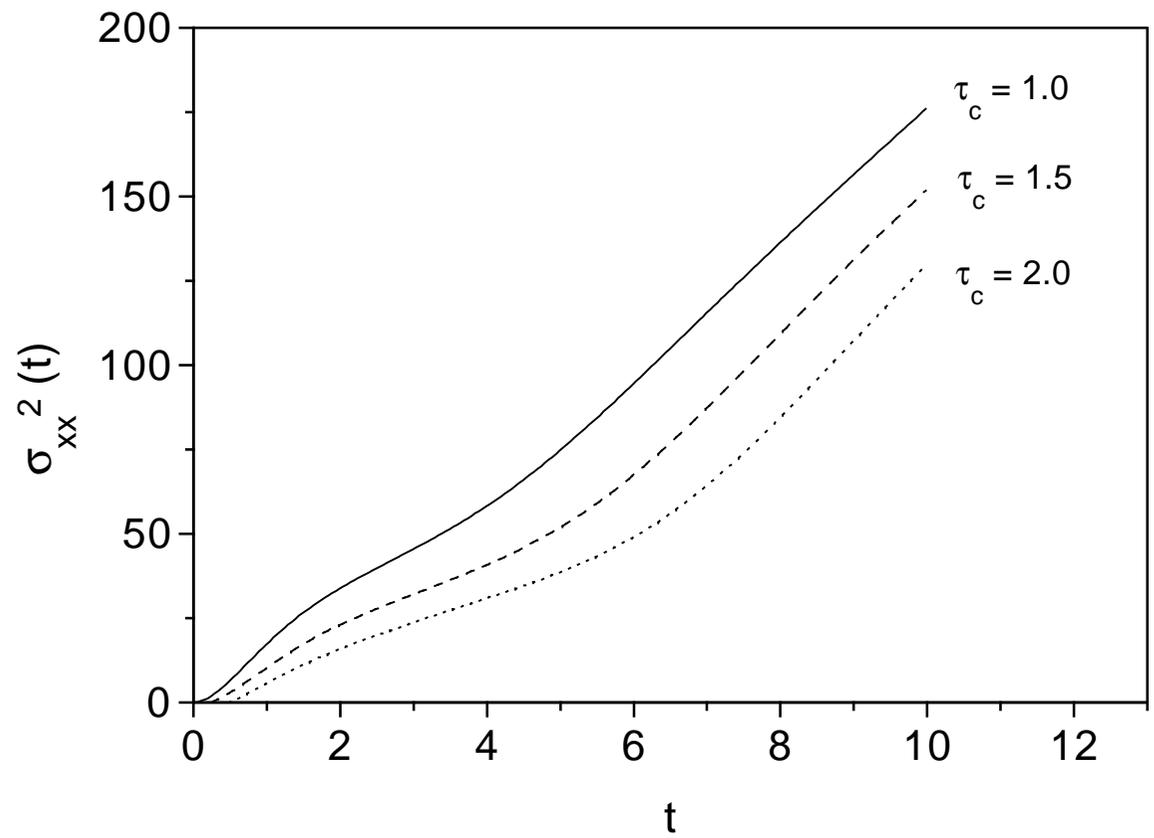
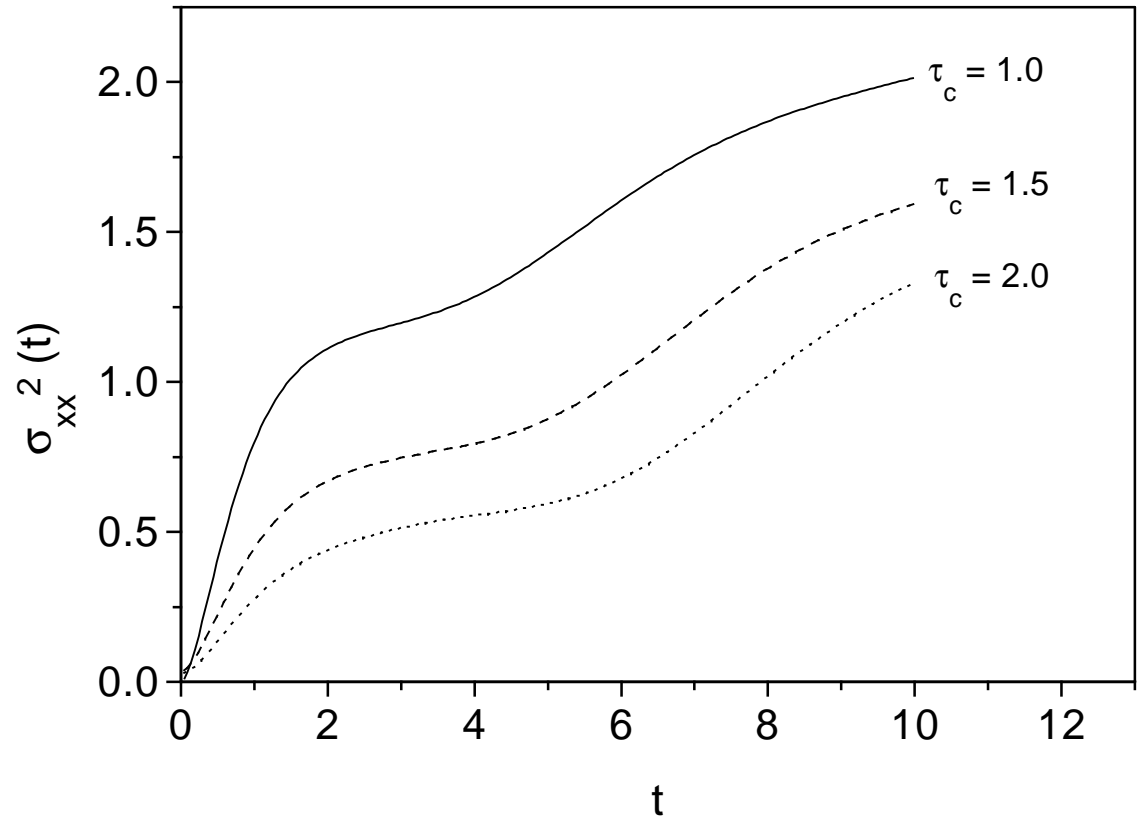
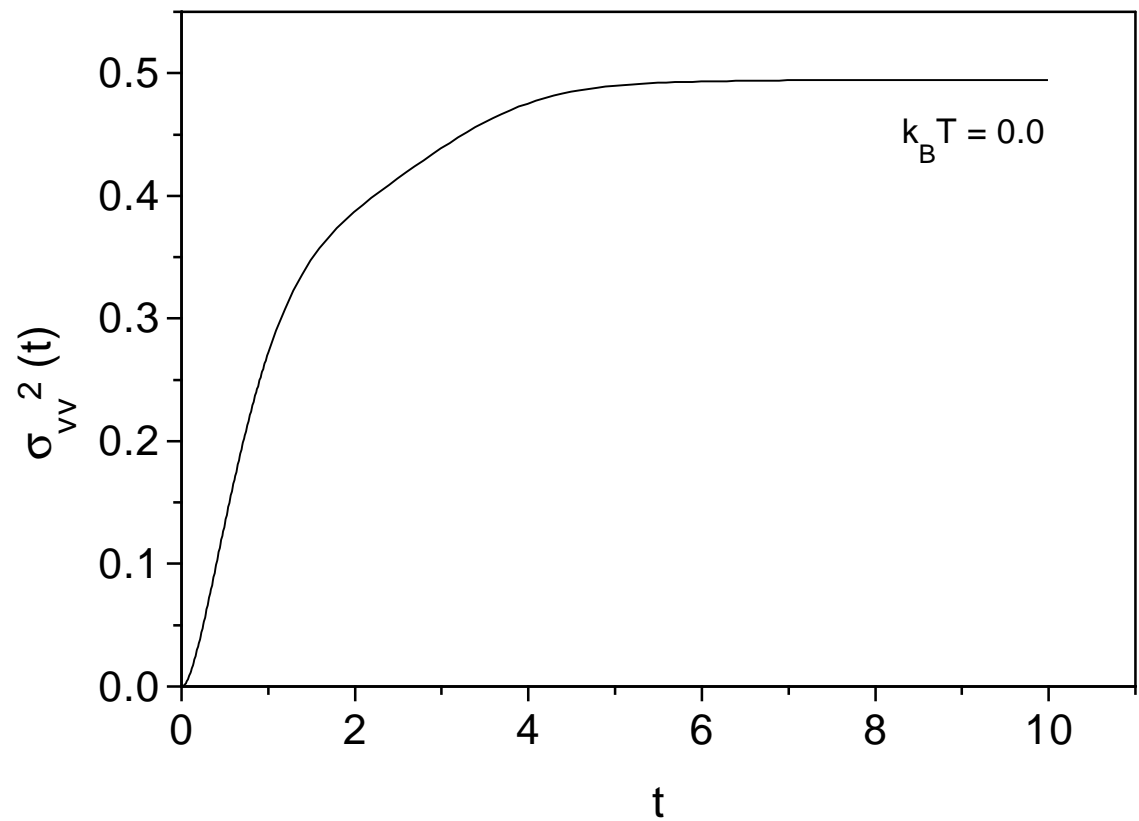
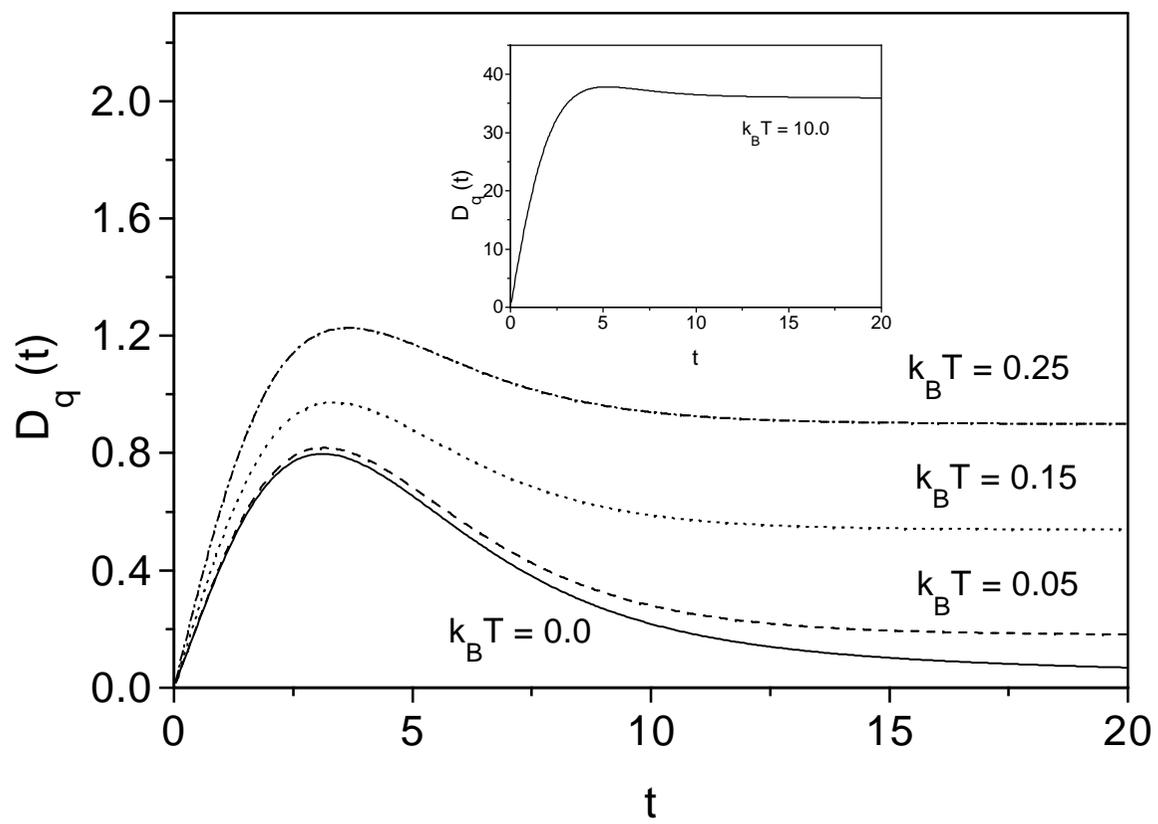
\end{document}